\documentclass[aps,prd,nofootinbib]{revtex4}

\begin{document}

\title{Schumann resonance transients and the search for gravitational waves}

\author{Z.K. Silagadze}
\email{Z.K.Silagadze@inp.nsk.su}
\affiliation{ Budker Institute of
Nuclear Physics and Novosibirsk State University, Novosibirsk 630
090, Russia }

\begin{abstract}
Schumann resonance transients which propagate around the globe can potentially 
generate a correlated background in widely separated gravitational wave 
detectors. We show that due to the distribution of lightning hotspots around 
the globe these transients have characteristic time lags, and this feature 
can be useful to further suppress such a background, especially  in searches 
of the stochastic gravitational-wave background. A brief review of the 
corresponding literature on Schumann resonances and lightnings  is also given. 
\end{abstract}

\maketitle

\section{Introduction}
The detection of gravitational-wave signals from inspiralling binary black 
holes by the Advanced Laser Interferometer Gravitational-Wave Observatory 
(LIGO) \cite{1,2,3} opens a new era in observational astrophysics and thus
is of paramount importance. Recent three-detector (including Virgo) 
observation of the gravitational wave GW170814 \cite{3A}, enabling a new class 
of phenomenological tests of gravity, as well as unprecedented joint 
gravitational and electromagnetic observation of a neutron star merger 
\cite{3B} are further solid confirmations that this new era indeed has come. 

As the gravitational-wave detectors are extremely sensitive instruments they 
are prone to many sources of noise that need to be identified and removed from 
the data. An impressive amount of efforts were undertaken by the LIGO 
collaboration to ensure that GW150914 signal was really the first detection 
of gravitational waves with all  transient noise backgrounds being under 
a good control \cite{4,5,6}.

Correlated magnetic fields from Schumann resonances constitute a well known 
potential source of correlated noise in gravitational waves detectors \cite{11,
12,13,13A,13B}. Such correlated noise may be a serious limiting factor in 
a search of stochastic gravitational-wave background which is expected to be 
created due to many independent, uncorrelated, not individually resolvable 
astrophysical and cosmological sources.

For short duration gravitational-wave transients, like the gravitational-wave 
signals observed by LIGO, Schumann resonances are not considered as significant
noise sources because the magnetic field amplitudes induced by even strong 
remote lightning strikes usually are of the order of a picotesla,  too small 
to produce strong signals in the LIGO gravitational-wave channel \cite{4}.

However, a recent study of short duration magnetic field transients, that were 
coincident in low-noise magnetometers in Poland and Colorado, revealed that 
there was about 2.3 coincident events per day where the amplitude of one of 
the pulses exceeded 200 pT \cite{21}. This finding indicates that  there would 
be a few such events per day that would appear simultaneously at the LIGO
gravitational-wave detector sites and could move the test masses by the 
amplitude that is comparable to the amplitude of the real GW150914 event 
\cite{21}.

Although it is highly unlikely that under the adopted data analysis strategy 
this relatively rare strong Schumann  resonance transients can imitate a real 
gravitational wave signal, they certainly deserve a further study.

\section{Schumann resonances}
Schumann resonances are global electromagnetic resonances in the 
Earth-ionosphere cavity \cite{14,15}. The electromagnetic waves in the 
extremely low frequencies (ELF) range (3Hz to 3 kHz) are mostly confined in 
this spherical cavity and their propagation is characterized by very low 
attenuation which in the 5 Hz to 60 Hz frequency range is of the order of 
0.5-1~db/Mm. Schumann resonances are eigenfrequencies of the Earth-ionosphere 
cavity. They are constantly excited by  lightning discharges around the globe. 
While individual lightning signals below 100 Hz are very weak, thanks to the 
very low attenuation, related ELF electromagnetic waves can be propagated 
a number of times around the globe, constructively interfere for wavelengths 
comparable with the Earth's circumference and create standing waves in 
the cavity.

For the ideal Earth-ionosphere cavity, Schumann obtained the following formula
for the resonance frequencies \cite{14,15}
\begin{equation}
f_n=\frac{c}{2\pi R}\,\sqrt{n(n+1)},
\label{eq1}
\end{equation}
where $R=6371\,\mathrm{km}$ is Earth's mean radius and $c$ is light velocity 
in vacuum. The observed Schumann resonance frequencies 7.8, 14.8, 19.8, 26~Hz
are lower than predicted by (\ref{eq1}) because the real Earth-ionosphere 
cavity is not an ideal one. Note that there exists some day-night variation of 
the resonance frequencies, and some catastrophic events, like a nuclear 
explosion, simultaneously lower all the resonance frequencies by about 0.5~Hz
due to lowering of the effective ionosphere height \cite{16}. Interestingly,
frequency decrease of comparable magnitude of the first Schumann resonance,
caused by the extremely intense cosmic gamma-ray flare, was reported in 
\cite{17}.

Usually eight distinct Schumann resonances are reliably detected in the 
frequency range from 7~Hz to 52~Hz. However five more were detected thanks
to particularly intense lightning discharges, thus extending the frequency 
range up to 90~Hz \cite{18}.

In fact (\ref{eq1}), as a formula describing oscillations in a spherical 
condenser with the small separation distance between spheres, was obtained
already in 1894 by Joseph Larmor \cite{19}. In general, Schumann resonances 
history is an interesting story including such names as George Fitzgerald 
and Nikola Tesla \cite{19}.

Phase velocity $V_p$ of the ELF electromagnetic wave is approximately given 
by the formula \cite{15} ($f$ is the frequency in Hz)
\begin{equation}
V_p=\frac{c}{1.64-0.179\ln{f}+0.0179\ln^2{f}},
\label{eq2}
\end{equation}
which for $f=10~\mathrm{Hz}$ gives $V_p=0.75\, c$.

Interestingly enough, the Schumann resonances make the Earth a natural 
gravitational-wave detector, albeit not very sensitive \cite{20}. As the Earth
is positively charged with respect to ionosphere, a static electric field,
the so-called fair weather field is present in the earth-ionosphere cavity.
In the presence of this background electric field, the infalling 
gravitational wave of suitable frequency resonantly excites the Schumann 
eigenmodes, most effectively the second Schumann resonance \cite{20}. 
Unfortunately, it is not practical to turn Earth into a gravitational-wave 
detector. Because of the weakness of the fair weather field (about 100~V/m)
and low value of the quality factor (from 2 to 6) of the Earth-ionosphere
resonant cavity, the sensitivity of such detector will be many orders of
magnitude smaller than the sensitivity of the modern gravitational-wave 
detectors. 

\section{ELF field transients}
The main source of the Schumann ELF waves are negative cloud-to-ground 
lightning discharges with the typical charge moment change of about 6~Ckm.
On Earth, storm cells, mostly in the tropics, generate about 50 such 
discharges per second. They generate the background Schumann resonance field 
with about $1\,\mathrm{pT}/\sqrt{\mathrm{Hz}}$ spectral density in the first 
Schumann resonance peak.

The positive cloud-to-ground atmospheric discharges have significantly higher
charge moment changes of about 250~Ckm \cite{21}. These strong lightnings
generate ELF field transients clearly  visible above the Schumann resonance 
background noise even at at large distances.

The so-called Q-bursts are more strong positive cloud-to-ground atmospheric 
discharges with charge moment changes of order of 1000~Ckm. ELF pulses excited
by Q-bursts propagate around the globe. At very far distances only the low 
frequency components of the ELF pulse will be clearly visible, because the 
higher frequency components experience more attenuation than the lower 
frequency components.

Transient luminous events, such as Sprites and Gigantic Jets, are associated 
to the most powerful discharges between the clouds and the ionosphere. Charge 
moment change in this case can reach several thousand Ckm \cite{21}. At that
the occurrence of a Sprite is usually preceded by a positive cloud-to-ground 
discharge.

Gigantic Jets generate ELF field pulses with very large amplitudes, visible 
anywhere on Earth. A Gigantic Jet near Corsica on December 12, 2009 generated
the transient magnetic field that was simultaneously observed as a very loud 
signal in magnetometers of the Virgo and LIGO-Hanford gravitational-wave 
detectors, and as a somewhat smaller signal in the LIGO-Livingston 
magnetometer. At that this Gigantic Jet actually created a perceptible 
gravitational-wave signal in Virgo detector (no signal was reliably observed 
in the LIGO detectors although their spectrograms also do show some 
disturbances that are coincident with the event) \cite{21}.

Fortunately Gigantic Jets are relatively rare, from a few to a dozen per year. 
However it was demonstrated in \cite{21} that relatively frequent magnetic 
transient events of sufficiently large amplitudes can be observed in 
coincidence on global distances. Therefore it can be expected that strong
ELF transients, related to powerful lightnings and Q-bursts across the globe, 
may be potential sources of correlated background for gravitational-wave 
detectors. 

\section{Time lags of ELF field transients for LIGO detectors}
In \cite{22} Earth's lightning hotspots are revealed in detail using 16 years 
of space-based Lightning Imaging Sensor observations. Information about
locations of these lightning hotspots allows us to calculate time lags 
between arrivals of the ELF transients from these locations to the 
LIGO-Livingston (latitude $30.563^\circ$, longitude $-90.774^\circ$) and
LIGO-Hanford (latitude $46.455^\circ$, longitude $-119.408^\circ$) 
gravitational-wave detectors.

ELF transients propagate along great circles paths and it is convenient to
calculate the relevant approximate distances by haversine formula \cite{23,24}
\begin{equation}
d=2R\arcsin{\left(\sqrt{\sin^2{\left(\frac{\varphi_2-\varphi_1}{2}\right)}+
\cos{\varphi_1}\cos{\varphi_2}\,\sin^2{\left(\frac{\lambda_2-\lambda_1}{2}
\right)}}\right )},
\label{eq3}
\end{equation}
where $\varphi_1,\,\varphi_2$ and $\lambda_1,\,\lambda_2$ are latitudes and
longitudes of two points on the Earth's surface, respectively.

We also need ELF transient propagation velocity. This was measured in 
\cite{25} using the round-the-world ELF transient signals and turned out to
be \cite{25,26} $V_g=265\pm 1 \mathrm{Mm/s}$ that is about 0.88c.
 
We have taken Earth's lightning hotspots from \cite{22} with lightning 
flash rate densities more than about $100\,\mathrm{fl}\; \mathrm{km}^{-2}
\mathrm{yr}^{-1}$ and calculated the expected time lags between ELF transients
arrivals from these locations to the LIGO detectors. The results are summarized
in the Table~\ref{table}.
\begin{table}[htp]
\caption{Top lightning hotspots with flash rate densities (FRD) more than
about $100\,\mathrm{fl}\; \mathrm{km}^{-2}\mathrm{yr}^{-1}$, their  latitude 
(Lat.) and longitude (Long.) positions, distances to the LIGO-Hanford (LHO) 
and LIGO-Livingston (LLO) detectors in km, and expected time lags in ms.}
{\begin{tabular}{| c | c | c | c | c | c |}
\hline
FRD/Region & Lat. & Long. & d to LHO &  d to LLO & Time lag \\ \hline
232.52/South America & 9.75 &  -71.65 & 6072 & 3045 & 11.5 \\ \hline
172.29/South America & 7.55 & -75.35 & 6018 & 3021 & 11.4 \\ \hline
138.61/South America & 8.85 & -73.05 & 6055 & 3035 & 11.4 \\ \hline
124.26/South America & 5.75 & -74.95 & 6206 & 3217 & 11.3 \\ \hline
114.19/South America & 8.45 &  -74.55 & 5991 & 2981 & 11.4 \\ \hline
105.73/South America & 8.15 & -76.85 & 5867 & 2882 & 11.3 \\ \hline\hline
205.31/Africa & -1.85 & 27.75 & 14123 & 12821  &  4.9 \\ \hline
176.71/Africa & -3.05 & 27.65 & 14235 & 12884 & 5.1  \\ \hline
143.21/Africa & -0.95 & 27.95 & 14044 & 12784 & 4.8  \\ \hline
129.58/Africa & 5.25  & 9.35  & 12375 & 10673 & 6.4  \\ \hline
129.50/Africa & 0.25  & 28.45  & 13951 & 12756 & 4.5  \\ \hline
127.52/Africa & -1.55 & 20.95  & 13717 & 12166 & 5.9  \\ \hline
117.98/Africa & 0.55  & 20.35  & 13483 & 11985 & 5.7  \\ \hline
117.19/Africa & -2.45 & 26.95  & 14139 & 12783 & 5.1  \\ \hline
116.78/Africa & 6.95  & 10.45  & 12290 & 10677 & 6.1  \\ \hline
112.17/Africa & 0.35  & 26.65  & 13849 & 12584 &  4.8 \\ \hline\hline
143.11/Asia & 34.45  & 72.35  & 10942 & 12573 & -6.2  \\ \hline
121.41/Asia & 33.35  & 74.55  & 11031 & 12743 & -6.5  \\ \hline
118.81/Asia & 33.75  & 70.75  & 11039 & 12605 & -5.9  \\ \hline
108.03/Asia & 14.55  & 43.45  & 13018 & 13007 &  0.0 \\ \hline
104.59/Asia & 33.85  & 73.25  & 10996 & 12659 &  -6.3 \\ \hline
101.79/Asia & 25.25  & 91.95  & 11440 & 13802 & -8.9  \\ \hline\hline
116.76/North America & 14.35  & -91.15  & 4433 & 1803 & 10.0  \\ \hline
103.23/North America & 14.85  & -92.05  & 4336 & 1752 &  9.8 \\ \hline
100.63/North America & 22.35  & -83.25  & 4156 & 1138 & 11.4 \\ \hline
100.24/North America & 18.55  & -74.35  & 5139 & 2127 & 11.4  \\ \hline
99.39/North America & 13.15  & -87.25  & 4767 & 1970 & 10.6  \\ \hline
\end{tabular}
\label{table} }
\end{table}
Note that the observed group velocity for short ELF field transients depends 
on the upper frequency limit of the receiver \cite{21}. For the magnetometers 
used in \cite{21} this frequency limit was 300~Hz corresponding to the quoted 
group velocity of about 0.88c. For the LIGO detectors the coupling of magnetic 
field to differential arm motion decreases by an order of magnitude for 30~Hz
compared to 10~Hz \cite{4}. Thus for the  LIGO detectors, as the ELF 
transients receivers, the more appropriate upper frequency limit is about 
30~H, not 300~Hz. According to (\ref{eq2}), low frequencies propagate with 
smaller velocities 0.75c-0.8c. Therefore the inferred time lags in the 
Table\ref{table} might be underestimated by about 15\%.

\section{Concluding remarks}
If the strong lightnings and Q-bursts indeed contribute to the LIGO detectors
correlated background when the distribution of lightning hotspots around the 
globe can lead to some regularities in this correlated noise. Namely, ELF 
transients due to lightnings in Africa will be characterized by 5-7~ms time 
lags between the LIGO-Hanford and LIGO-Livingston detectors. Asian lightnings 
lead to time lags which have about the same magnitude but the opposite sign. 
Lightnings in North and South Americas should lead positive time lags of about
11-13~ms, greater than the light propagation time between the LIGO-Hanford and 
LIGO-Livingston detectors.

We don't think that such correlated background represent a serious obstacle
in the gravitational-wave search because really strong lightnings and Q-bursts 
are relatively rare. However, as was mentioned in \cite{21}, it will be useful 
to have a carefully designed low-noise magnetometers system, installed at 
electromagnetically quiet locations, to further mitigate such a background 
noise. Maybe some of such ELF transient events can be used even for 
calibration purposes. Namely to monitor the coupling of magnetic field to 
differential arm motion in the gravitational-wave detectors.  

LIGO and Virgo magnetometers routinely detect Schumann resonance background
signal (see, for example, \cite{13A}). The observation of individual ELF 
transients is, however, more complicated task because the magnetometers 
operate in the quite noisy  magnetic environments at the LIGO and Virgo sites.
Therefore, the above mentioned additional system of magnetometers will be very 
helpful if we wish to reliable detect all relevant global ELF transients.
Additionally, this global low-noise magnetometers system can be used in the
lightning studies which is by itself an interesting and important scientific
topic \cite{14}.

It is usually assumed that in the far field region the horizontal 
electric-field and vertical magnetic field tend to zero. Recently, it was 
demonstrated that in reality this assumption is not always justified probably 
because of local inhomogeneities of the ground's conductivity \cite{28}. As 
a result some of the magnetic field's energy can hide in the vertical 
component. This circumstance should be taken into consideration when designing
ELF transients monitoring system.

Observations indicate that the background and transient activity in Schumann 
Resonances are linked. Namely, the intensity of the background is roughly
proportional to the number of large mesoscale lightnings \cite{29}. The 
mechanism of this connection is the following \cite{29,30}. The background 
signal represents the result of integration of lightnings from the ordinary 
late-afternoon thunderstorms at lightning hotspots. Later these thunderstorms 
amalgamate into late evening mesoscale convective systems which are good
producers of mesoscale lightning flashes with exceptionally energetic 
transients. However all details of this interaction are still unknown and 
deserve a thorough study \cite{30}. 
 
The suppression of the Schumann resonances related background is especially
important in searches of stochastic gravitational-wave background. We hope 
that the fact that the Schumann resonance transients are expected to have 
characteristic time lags will help to reduce this background. 

A special ELF transients monitoring system, if developed, will allow to 
investigate the above mentioned fundamental interaction between background 
and transient phenomena in Schumann resonances and the acquired  knowledge
can be used to reduce the corresponding correlated background noise in
stochastic gravitational-wave background studies. It seems natural for 
gravitational-wave and lightning communities to unify their efforts for mutual
benefit.  

\section*{Acknowledgments}
The work is supported by the Ministry of Education and Science of the Russian 
Federation. 


\end{document}